\documentclass[preprint,aps,12pt,preprintnumbers,eqsecnum,nofootinbib,
superscriptaddress]{revtex4}
\usepackage{graphicx}
\usepackage{amssymb,amsmath}

\def\be{\begin{equation}}
\def\ee{\end{equation}}
\def\beq{\begin{eqnarray}}
\def\eeq{\end{eqnarray}}

\begin{document}
%
%
\title{\vspace*{0.5in} Tribimaximal Neutrino Mixing from $A_4$
Replication}
\vskip 0.1in
\author{Christopher D. Carone}\email[]{cdcaro@wm.edu}
\affiliation{Particle Theory Group, Department of Physics,
College of William and Mary, Williamsburg, VA 23187-8795}
\author{Richard F. Lebed}\email[]{Richard.Lebed@asu.edu}
\affiliation{Department of Physics, Arizona State University, Tempe,
AZ 85287-1504}
\date{November 2010}
\begin{abstract}
Motivated by dimensional deconstruction, we propose a model of
tribimaximal neutrino mixing based on $A_4\times A_4$ symmetry.  In
this model, the two triplet symmetry-breaking fields of conventional
$A_4$ models are taken to transform under different $A_4$ group
factors, but are not distinguished by any other quantum numbers.  An
additional bi-triplet flavon field breaks $A_4 \times A_4$ to its
diagonal subgroup.  If the bi-triplet transforms under an additional
$Z_3$ symmetry, we show that one can construct a general,
renormalizable superpotential that yields the desired pattern of
symmetry breaking.  We identify the features that this model has in
common with a deconstructed 5D theory in which $A_4$ is a subgroup of
a continuous gauged flavor symmetry in the bulk.
\end{abstract}
\maketitle

\section{Introduction}

The observed pattern of fermion masses and mixing angles remains one
of the major unresolved mysteries of the standard model.  In the
neutral sector, a global analysis of the neutrino oscillation data
leads to the mixing angles~\cite{Aharmim:2009gd}
\begin{equation}
\sin^2\theta_{12}=0.314^{+0.019}_{-0.014} \, ,  \,\,\,\,\,
\sin^2\theta_{23}=0.50^{+0.07}_{-0.06} \, ,     \,\,\,\,\,
\sin^2\theta_{13}<0.057  \, ,
\label{eq:nuangles}
\end{equation}
where the limit on $\sin^2\theta_{13}$ corresponds to the 95\%
confidence level.  Unlike the mixing angles of the
Cabibbo-Kobayashi-Maskawa (CKM) matrix, the nonzero angles in
Eq.~(\ref{eq:nuangles}) are large.  Even more intriguing is that these
angles are consistent with the values
\begin{equation}
\sin^2\theta_{12}= \frac 1 3 \, ,  \,\,\,\,\,
\sin^2\theta_{23}= \frac 1 2 \, ,     \,\,\,\,\,
\sin^2\theta_{13}= 0   \, ,
\label{eq:tbmangles}
\end{equation}
which may be obtained from a rotation matrix of tribimaximal (TB)
form~\cite{TB}
\begin{equation} \label{UTB}
U_{\rm TB} = \left(
\begin{array}{ccc}
 \sqrt{\frac{\textstyle 2}{\textstyle 3}} &
\frac{\textstyle 1}{\sqrt{\textstyle 3}} & 0 \\
-\frac{\textstyle 1}{\sqrt{\textstyle 6}} &
\frac{\textstyle 1}{\sqrt{\textstyle 3}} &
-\frac{\textstyle 1}{\sqrt{\textstyle 2}} \\
-\frac{\textstyle 1}{\sqrt{\textstyle 6}} &
\frac{\textstyle 1}{\sqrt{\textstyle 3}} &
+\frac{\textstyle 1}{\sqrt{\textstyle 2}}
\end{array}
\right) \, .
\end{equation}
Here, for simplicity, we have omitted possible phases.  Models that
lead to the mixing matrix in Eq.~(\ref{UTB}) are phenomenologically
relevant, and numerous examples have been proposed over the past few
years.  (For reviews of the substantial literature,
see~\cite{nureviews}.)

Most models that aim to explain TB neutrino mixing from an underlying
symmetry principle are based on the discrete group
$A_4$~\cite{a41,a42,a43,a45,a46,a4review,a44,Csaki:2008qq,Ma:2009wi}.
In these models, the $A_4$ symmetry is broken at lowest order to a
$Z_2$ subgroup in the neutral-lepton sector, and to a $Z_3$ subgroup
in the charged-lepton sector; this breaking is accomplished by the
vacuum expectation values (vevs) of two triplet fields, denoted
$\phi_S$ and $\phi_T$, respectively. The necessity of fields that
prefer different symmetry-breaking directions and that must couple
non-generically to the lepton fields suggests that physics other than
$A_4$ symmetry is necessary to account for the TB mass matrix
textures.  (For the alternative viewpoint see, for example,
Ref.~\cite{Ma:2009wi}; for models that employ
a different vev structure, see Ref.~\cite{fortheref}.) Possible
strategies include, for example, enlarging the flavor symmetry group
by including additional Abelian factors, or exploiting a
higher-dimensional construction in which the localization of the
fields at specific points in an extra-dimensional interval account for
a non-generic set of couplings in the four-dimensional (4D) theory, as
in Refs.~\cite{a44,Csaki:2008qq}. In this letter, we consider a
different class of 4D theories, those that have an enlarged flavor
symmetry group, but are constructed to mimic closely the effects of
localization in five-dimensional (5D) theories.

The deconstruction of 5D theories~\cite{decon} has led in the past to
new and interesting 4D models.  Applications of deconstruction to
models of electroweak symmetry breaking~\cite{deconew} and
supersymmetry breaking~\cite{deconsusy} are well known.  Although the
resulting models have a potentially large, replicated group structure,
these theories have garnered substantial attention since they inherit
some of the novel features of their 5D progenitors, even when the
number of lattice sites is taken to be small.  The application of this
approach to $A_4$ theories has not been considered before and
motivates the structure of our model: The flavor group is $A_4 \times
A_4$ and is broken to the diagonal $A_4$ by a bi-triplet field
$\Sigma$ that transforms as a $({\bf 3},{\bf 3})$.  As we discuss in
Sec.~\ref{sec:5d}, this structure is exactly as expected from the
two-site deconstruction of a 5D theory in which a continuous gauged
flavor symmetry in the bulk is broken to a discrete subgroup.  The
triplets $\phi_S$ and $\phi_T$ of conventional $A_4$ models here
transform under different $A_4$ factors, corresponding to
extra-dimensional localization, allowing one to obtain the desired
pattern of couplings for the charged and neutral leptons.  Exploring
theories of this type is valuable because higher-dimensional
realizations of the flavor structure of the standard model can be
mapped into the simpler 4D theories.  For example, Yukawa suppression
factors originating from the localization of flavor symmetry-breaking
fields at different points in an extra dimension might be important
ingredients in a successful $A_4$ flavor model for quarks and leptons
in 5D, but also may be encoded in viable deconstructed models that
require only a small number of replicated $A_4$ factors.  However,
before pursuing such a path, one must know how to construct the
simplest theory of this type; hence we focus on a two-site model of
the lepton sector in the present work.  In particular, we obtain an
explicit, renormalizable superpotential that achieves the desired
flavor symmetry breaking and find charge assignments for the lepton
fields that lead to the desired phenomenology.

The paper is organized as follows.  In Sec.~\ref{sec:symbrk} we
construct an explicit, renormalizable superpotential involving
$\phi_S$, $\phi_T$, and $\Sigma$ such that $\phi_T$ breaks the first
$A_4$ factor to $Z_3$, $\phi_S$ breaks the second $A_4$ factor to
$Z_2$, and $\Sigma$ breaks $A_4 \times A_4$ to its diagonal subgroup;
the collective effect is to break the flavor group completely.  In
Sec.~\ref{sec:model} we show how this symmetry-breaking sector can be
used to construct a model of TB neutrino mixing.  In Sec.~\ref{sec:5d}
we discuss the relationship between the model and a deconstructed 5D
theory in which the $A_4$ triplet fields are localized.
Section~\ref{sec:conc} summarizes our conclusions and suggests
directions for future work.

\section{Symmetry Breaking} \label{sec:symbrk}

In this section we consider the sector of the model that spontaneously
breaks $A_4 \times A_4$.  The Clebsch-Gordan matrices for combining
$A_4$ representation are the same as those for combining the
odd-dimensional representations of $T^\prime$; our basis is defined by
the Clebsch-Gordan matrices given in the appendix of Ref.~\cite{acl2}, and
our notation for the one-dimensional representations is specified therein.
Below we denote the Clebsch-Gordan matrices that combine two triplets
into a ${\bf 1}_{0,\pm}$ singlet by $C_{0,\pm}$, and into the $k^{th}$
component of another triplet by $C_A^k$ or $C_S^k$, where $A(S)$
indicates the symmetry of the product under the interchange of the two
triplets.  A concise review for those unfamiliar with the group theory of
$A_4$ can be found in Ref.~\cite{a4review}.

Let us first consider the field $\Sigma$ that transforms as a $({\bf
3},{\bf 3})$ under $A_4 \times A_4$.  The vev
\begin{equation} \label{eq:desired}
\langle \Sigma \rangle = \left( \begin{array}{ccc} v & 0 & 0 \\
0 & 0 & v \\
0 & v & 0 \end{array}\right) \equiv v \, C_0
\end{equation}
breaks the flavor group to its diagonal $A_4$.  To see this, imagine
coupling the $\Sigma$ field to two triplets, $\phi_1 \sim ({\bf 3},
{\bf 1}_0)$ and $\phi_2 \sim ({\bf 1}_0, {\bf 3})$:
\begin{equation}
\phi_1^T C_0 \Sigma C_0 \phi_2 \, .
\end{equation}
The part proportional to the vev can be written
\begin{equation}
v\, \phi_1^T C_0^3 \phi_2 = v\, \phi_1^T C_0 \phi_2 \, ,
\end{equation}
which is a trivial singlet (${\bf 1}_0$) under the $A_4$ subgroup
under which both $\phi_1$ and $\phi_2$ transform identically.  The
individual $A_4$ factors are broken by a $({\bf 3}, {\bf 1}_0)$
triplet $\phi_T$ with the vev
\begin{equation}
\langle \phi_T \rangle = (v_T,0,0) \,\,\,,
\end{equation}
which breaks the first $A_4$ to a $Z_3$ subgroup, and a $({\bf 1}_0,
{\bf 3})$ triplet $\phi_S$ with the vev
\begin{equation}
\langle \phi_S \rangle = (v_S,v_S,v_S) \, ,
\end{equation}
which breaks the second $A_4$ to a $Z_2$ subgroup.  These vev patterns
are the ones required in conventional $A_4$ models of TB mixing.  Note
that the combined effect of the $\phi_T$, $\phi_S$, and $\Sigma$ vevs
is to break $A_4 \times A_4$ completely.

We now construct a supersymmetric potential that forces the vevs
described above, while preserving supersymmetry and avoiding
massless states.  It is convenient to define the following quadratic
combinations of $\Sigma$, each of which transforms as (${\bf 3},{\bf
3}$):
\begin{equation}
(\Sigma^2_{SS})^{i\alpha} \equiv \mbox{Tr } C_S^i \Sigma C_S^\alpha
\Sigma^T \,,
\end{equation}
\begin{equation}
(\Sigma^2_{AA})^{i\alpha} \equiv \mbox{Tr } C_A^i \Sigma C_A^\alpha
\Sigma^T \,.
\end{equation}
The $AS$ and $SA$ combinations are vanishing. It is also useful to
note that one can compose the following $({\bf 1}_0, {\bf 1}_0)$
invariants given a generic $A_4$ triplet $\phi$ with components
$(A,B,C)$ in our basis:
\begin{equation}
\phi^2 = A^2 + 2 B \, C  \,,
\end{equation}
\begin{equation}
\phi^3 = 2\, (A^3 + B^3 + C^3 - 3 A B C)  \,.
\end{equation}
Our superpotential is
\begin{equation}
W=W_T + W_S + W_\Sigma \,,
\end{equation}
where
\begin{equation}
W_T = m_0 \phi_T^2 + \lambda_T \phi_T^3+C_{st} S_0 \phi_T^2 \,,
\label{eq:wtsp}
\end{equation}
\begin{equation}
W_S = m_{S1}^2 \, S_0 + m_{S2} \, S_0^2 +\lambda_S \, S_0^3 + \kappa_1
(S_0-m_{S3}) \, \phi_S^2 + \lambda \, \phi_S^3 \,,
\label{eq:wssp}
\end{equation}
\begin{eqnarray} \label{eq:wsigsp}
W_\Sigma &=& x_1 \mbox{Tr } \Sigma^2_{SS} C_0 \Sigma^T C_0+x_2
\mbox{Tr } \Sigma^2_{AA} C_0 \Sigma^T C_0
+x_3 \, S_+ \, \mbox{Tr } \Sigma C_0 \Sigma^T C_0  \nonumber \\
&+&\kappa_2 (S_0 - m_\pm)\, S_+ \,S_- + \lambda_1 S_+^3 +
\lambda_2 S_-^3 \,.
\end{eqnarray}
Here $C_{st}$, $\lambda_S$, $\lambda_T$, $\kappa_{1,2}$, $\lambda$,
$\lambda_{1,2}$, and $x_{1,2,3}$ are couplings constants.  One can
show that renormalizable couplings between $\Sigma$ and $\phi_S$ lead
to the spontaneous breaking of supersymmetry, and we therefore forbid
such terms by imposing a $Z_3$ symmetry under which $\Sigma
\rightarrow \omega \Sigma$, where $\omega^3=1$.  Henceforth, the
charges $0$, $+$, and $-$ refer to fields that transform by $1$,
$\omega$, and $\omega^2$ under the $Z_3$ factor, respectively.  The
fields $S_0$, $S_+$, and $S_-$ are $A_4 \times A_4$ singlets with
$Z_3$ charges $0$, $+$, and $-$, respectively.  The superpotential
given by Eqs.~(\ref{eq:wtsp})--(\ref{eq:wsigsp}) is the most general
one consistent with the $A_4 \times A_4 \times Z_3$ symmetry.  We note
that it is possible to construct models in which this $Z_3$ is not a
flavor symmetry, in the sense that it does not act on the lepton
fields; however, we find it convenient to include the $Z_3$ as part of
the flavor group, which aids us later in obtaining an appropriate
charged-lepton mass hierarchy.

To study the F-flatness conditions, we find it convenient to
parametrize the components of the various $A_4 \times A_4$ multiplets
as follows:
\begin{equation}
\Sigma = \left[\begin{array}{ccc}
a+2 \,d & b-e-h & c-f+j \\
b-e+h & c+2\, f& a-d-g \\
c-f-j & a-d+g & b+2 \,e \end{array}\right]  \,,
\end{equation}
\begin{equation}
\phi_T = (a_1,b_1,c_1) \,,
\end{equation}
\begin{equation}
\phi_S = (a_2,b_2,c_2) \,.
\end{equation}
Including the singlets $S_0$, $S_+$, and $S_-$, one has 18 F-flatness
conditions.  Most of them vanish automatically for our assumed pattern
of vevs, except for $F_a$, $F_{a_1}$, $F_{a_2}$, $F_{b_2}$, $F_{c_2}$,
$F_{S_0}$, $F_{S_+}$, and $F_{S_-}$.  Note that $F_{a_2}$, $F_{b_2}$,
and $F_{c_2}$ all vanish if
\begin{equation}
\langle S_0 \rangle \equiv v_0 = m_{S3} \, .
\end{equation}
One is left with five remaining equations and five unknowns: $\langle
a \rangle\equiv v$, $\langle a_1 \rangle \equiv v_T$, $\langle a_2
\rangle =\langle b_2 \rangle=\langle c_2 \rangle \equiv v_S$, $v_+
\equiv \langle S_+ \rangle$, and $v_- \equiv \langle S_- \rangle$,
which can be solved analytically; a solution with the desired
nonvanishing vevs is given by:
\begin{equation}
v_T = -\frac{1}{3\lambda_T} (m_0 + C_{st} \, m_{S3}) \,,
\end{equation}
\begin{equation}
v_+ = - \frac{\kappa_2 (m_{S3}-m_{\pm})}
{3 \lambda_2^{1/3}}\left[
\frac{x_3^3}{9\,(3 \,x_1+x_2)^2}+ \lambda_1\right]^{-2/3} \,,
\end{equation}
\begin{equation}
v_- = -\frac{\kappa_2 (m_{S3}-m_{\pm})} {3 \lambda_2^{2/3}}\left[
\frac{x_3^3}{9\,(3\, x_1+x_2)^2}+ \lambda_1\right]^{-1/3} \,,
\end{equation}
\begin{equation}
v = -\frac{x_3}{3\,(3\,x_1+x_2)} \, v_+ \, ,
\end{equation}
\begin{eqnarray}
\hspace{-1.5em}
v_S &=& \frac{1}{(3 \kappa_1)^{1/2}} \left[
-\frac{C_{st}}{9 \lambda_T^2} (m_0 +C_{st} m_{S3})^2
-m_{S1}^2-2 m_{S2} m_{S3}
- 3 \lambda_S m_{S3}^2 - \kappa_2 v_+ v_- \right]^{1/2} \! \! .
\end{eqnarray}
This solution represents a distinct point in the moduli space of the
theory, and it is straightforward to verify directly that the scalar
potential at this point has a positive-definite second-derivative
matrix.  The mass scale of the scalar states is set by the various
dimensionful parameters found in
Eqs.~(\ref{eq:wtsp})--(\ref{eq:wsigsp}), and is much larger than the
weak scale (see the following section).  Since soft
supersymmetry-breaking effects are associated with a much lower scale,
they will at most perturb this solution, but might be chosen to render
it a global minimum compared to other possible supersymmetric vacua.
Alternatively, the desired symmetry breaking might be achieved at a
local minimum of the potential, which is sufficient for our present
purposes.

\section{The Model} \label{sec:model}

In the previous section, we constructed a superpotential leading to
the vevs $\langle \phi_S \rangle = (v_S,v_S,v_S)$, $\langle \phi_T
\rangle = (v_T,0,0)$ and $\langle \Sigma \rangle = v \, C_0$.  The
superpotential is the most general one consistent with the discrete
symmetry $A_4 \times A_4 \times Z_3$, where the $Z_3$ factor acts on
the $\Sigma$ field, but not on the triplets $\phi_S$ and $\phi_T$.
Now we consider the flavor structure of the lepton sector that follows
from this symmetry breaking and show that TB neutrino mixing results.
We include higher-dimensional operators that contribute to the lepton
mass matrices, suppressed by powers of a flavor scale $\Lambda$; these
operators may have a renormalizable origin, with $\Lambda$ identified
as the mass scale of heavy, vector-like states~\cite{fn}.

We assume the following $A_4 \times A_4 \times Z_3$ charge assignment
for the lepton fields:
\begin{eqnarray}
 & \nu \sim ({\bf 1}_0,{\bf 3})_0\, , \,\,\,\,\, L \sim
({\bf 3},{\bf 1}_0)_-\,, & \nonumber \\
 & E_1 \sim ({\bf 1}_0, {\bf 1_+})_+ \,,\,\,\,\,\, E_2 \sim
({\bf 1_-}, {\bf 1}_0)_- \,, \,\,\,\,\,
E_3 \sim ({\bf 1_+}, {\bf 1}_0)_+ &\,.
\end{eqnarray}
All the fields shown are left-handed chiral superfields, and the final
subscript indicates the $Z_3$ charge.  In this notation, the flavon
fields transform as $\phi_T \sim ({\bf 3}, {\bf 1}_0)_0$, $\phi_S \sim
({\bf 1}_0, {\bf 3})_0$, $\Sigma \sim ({\bf 3}, {\bf 3})_+$, $S_0 \sim
({\bf 1}_0, {\bf 1}_0)_0$, $S_+ \sim ({\bf 1}_0, {\bf 1}_0)_+$, and
$S_- \sim ({\bf 1}_0, {\bf 1}_0)_-$.  Crucially, the right-handed
neutrino $\nu$ and the triplet flavon $\phi_S$ transform under a
different $A_4$ factor than the $L$ and $E_{2,3}$ superfields.  Hence,
at lowest order the Majorana mass matrix for the right-handed
neutrinos originates from
\begin{equation}
W_{RR} = M \nu \nu + c_1 \phi_S \, \nu\nu\, + \mbox{ higher order} ,
\label{eq:wrr}
\end{equation}
and does not involve $\phi_T$ in the renormalizable couplings.  Here,
$c_1$ is an undetermined coupling; we use similar notation in the
discussion that follows.  Equation~(\ref{eq:wrr}) leads to the
Majorana mass matrix
\begin{equation}
M_{RR} = \left(
\begin{array}{ccc}
M+2 c_1 v_S & - c_1 v_S & - c_1 v_S \\
- c_1 v_S & 2 c_1 v_S & M-c_1 v_S \\
- c_1 v_S & M-c_1 v_S & 2 c_1 v_S
\end{array} \right)  \,.
\label{eq:mrr}
\end{equation}
On the other hand, the neutrino Dirac mass matrix connects the $L$ and
$\nu$ superfields, which transform nontrivially under different $A_4$
factors.  At lowest order, precisely one operator contributes:
\begin{equation}
W_{LR} = \frac{c_2}{\Lambda} L H_U \Sigma \nu + \mbox{ higher order}\, ,
\end{equation}
where $H_U$ is the up-sector Higgs superfield.  The $\Sigma$ vev of
Eq.~(\ref{eq:desired}) yields the Dirac mass matrix
\begin{equation}
M_{LR}= \frac{c_2 \,v}{\Lambda} \langle H_U \rangle \left(
\begin{array}{ccc}
1 & 0 & 0 \\
0 & 0 & 1 \\
0 & 1 & 0
\end{array} \right) \,.
\label{eq:mlr}
\end{equation}
It is well known that the textures in Eqs.~(\ref{eq:mrr}) and
(\ref{eq:mlr}) lead to exact TB mixing, $\sin^2\theta_{12}= \frac 1
3$, $\sin^2 \theta_{23}= \frac 1 2$ and $\sin^2\theta_{13}=0$, if the
charged lepton mass matrix is diagonal~\cite{a4review}.  The neutrino
mass-squared ratio is given by
\begin{equation}
\frac{\Delta m_{21}^2}{\Delta m_{32}^2}= \frac{(M-3 c_1 v_S)^2}
{(M+3 c_1 v_S)^2} \cdot \frac{2 M+3 c_1 v_S}{2 M-3 c_1 v_S}\, ,
\end{equation}
which can accommodate the experimental value ($\approx 0.03$) with a
mild tuning of $M$ and $c_1 v_S$.

The $L$ and $E_{2,3}$ superfields transform under a single $A_4$
factor, and carry the charge assignments of conventional $A_4$ models,
aside from the additional $Z_3$ charge.  The differing $Z_3$ charges
of $E_2$ and $E_3$ lead automatically to $m_\mu \ll m_\tau$.  The $L$
and $E_1$ fields are charged under different $A_4$ factors, so that
couplings of these fields occur at even higher order.  In this way, a
hierarchy of charged-lepton masses is obtained.  The diagonal entries
of the charged-lepton Yukawa matrix receive leading-order
contributions from the operators
\begin{equation}
W_L = \frac{c_3}{\Lambda} L H_D E_3 \phi_T + \frac{c_4}{\Lambda^2}
L H_D E_2 \phi_T S_-
+ \frac{c_5}{\Lambda^3} L H_D E_1 \phi_T \phi_S^2
+\frac{c_6}{\Lambda^3} L H_D E_1 \Sigma \phi_S S_-
+ \cdots \,,
\end{equation}
where the first, second, and third terms contribute only to $L_3 E_3$,
$L_2 E_2$, and $L_1 E_1$, respectively.  Setting the flavon and Higgs
fields to their vevs, the diagonal entries of the mass matrix are
determined by
\begin{equation}
W_M = \frac{c_3 v_T}{\Lambda}   L_3 \langle H_D \rangle E_3
+\frac{c_4 v_- v_T}{\Lambda^2}   L_2 \langle H_D \rangle E_2
+\frac{3 c_5 v_S^2 v_T + c_6 v v_S v_-}{\Lambda^3}  L_1 \langle H_D
\rangle E_1 + \cdots \, .
\end{equation}
Possible operators contributing to $L_{1,2} E_3$ are higher order than
the one contributing to $L_3 E_3$, indicating that the $1$-$3$ and
$2$-$3$ rotations on $L$ needed to diagonalize the charged-lepton mass
matrix are small.  Similarly, the operators contributing to $L_1 E_2$
are higher order than the one contributing to $L_2 E_2$, indicating
that the $1$-$2$ rotation on $L$ needed to diagonalize the charged
lepton mass matrix is also suppressed.  Hence, TB neutrino mixing
remains as the leading-order prediction of the model.  Note that the
correct charged-lepton Yukawa couplings may be obtained by choosing
$v_T/\Lambda$, $v_S/\Lambda$, and $v_- /\Lambda = {\cal
O}(\lambda_C^2)$, where $\lambda_C \approx 0.22$ is the Cabibbo
angle. Assuming that all the flavor symmetry-breaking vevs are of
comparable size, it follows that the mixing angles required to
diagonalize the charged-lepton Yukawa matrix are at most ${\cal
O}(\lambda_C^2$).

Corrections of this size to the lowest-order TB mixing angles are
comparable to those originating from the higher-order corrections to
Eqs.~(\ref{eq:mrr}) and (\ref{eq:mlr}).  Deviations from exact TB
mixing can be described in terms of deviation
parameters~\cite{devoth,cldev}, such as $\tilde{\epsilon}$,
$\tilde{\delta}_1$, and $\tilde{\delta}_2$ defined in Eq.~(10) of
Ref.~\cite{cldev}. We find in the present model that
\begin{equation}
\tilde{\epsilon}\,,\,\,\,\, \tilde{\delta}_1\,, \mbox{ and }
\tilde{\delta}_2 = {\cal O}(\lambda_C^2) \,,
\end{equation}
which is comparable~\cite{cldev} to the experimental uncertainty on
these parameters.  Since the model's predictions depend on a number of
unknown ${\cal O}(1)$ operator coefficients, we find that the
experimental bounds on $\tilde{\epsilon}$, $\tilde{\delta}_1$, and
$\tilde{\delta}_2$ can easily be satisfied.

Finally, we comment on the origin of the scale $\Lambda$, which is
around $10^{13}$~GeV by the seesaw formula if one takes $M \sim v_S
\sim \lambda_C^2 \Lambda$ in Eq.~(\ref{eq:mrr}). All of the required
higher-dimensional operators may be generated by integrating out
heavy, vector-like matter fields.  As one example, the neutrino
Dirac-mass operator $L H_U \Sigma \nu / \Lambda$ may arise if a
vector-like lepton doublet with flavor quantum numbers
\begin{equation}
L^H \sim ({\bf 1}_0, {\bf 3})_0  \,\,\,\, \mbox{ and } \,\,\,\
\bar{L}^H \sim ({\bf 1}_0, {\bf 3})_0 \,
\end{equation}
is present in the ultraviolet (UV) completion of the effective theory.
Here, $L^H$ has the same gauge quantum numbers as $L$, while $L^H$ has
the conjugate values.  The gauge- and flavor-invariant superpotential
terms
\begin{equation}
W_{UV} = L \Sigma \bar{L}^H + L^H H_U \nu + M_H \bar{L}^H L^H \,
\end{equation}
lead to the desired dimension-four operator in the low-energy
effective theory, after identifying $\Lambda$ with $M_H$.  The
complete spectrum of vector-like states in the UV completion can be
found by starting with the desired higher-dimensional operators and
reverse-engineering in this way\footnote{Since the vector-like states
are assumed to have the same R-parity as the standard-model matter
fields, we do not generate any higher-dimensional operators that
affect the form of the superpotential discussed in Sec.~\ref{sec:symbrk}.}.

\section{Relationship to 5D Models}\label{sec:5d}

In the literature on TB neutrino mixing, models with a single $A_4$
symmetry have been proposed that use the localization of fields in an
extra dimension to forbid undesired couplings.  Although the model
proposed here is four dimensional, the structure of the model bears
some similarities to these higher-dimensional theories.  In this
section, we discuss the extent of these similarities.

It is useful to consider a simple example.  Imagine a 5D theory with a
gauged U(1) flavor symmetry in the bulk, as well as a bulk field
$\chi$ that has charge $Q$.  For definiteness, we assume that the U(1)
is normalized so that some other field in the theory has charge $+1$,
and that all the other U(1) charges are integral.  If the $\chi$ field
develops a nonvanishing profile $\langle \chi(y) \rangle$, where $y$
is the extra-dimensional coordinate, then the 5D continuous gauge
symmetry is broken to a $Z_Q$ subgroup, since
\begin{equation}
e^{i \, \alpha(x^\mu,y) \,Q} \langle \chi \rangle = \langle \chi
\rangle
\end{equation}
for $\alpha(x^\mu,y)=2 \pi /Q$.  Denoting $\exp(2\pi i/Q) =
\omega$, a field $\psi$ with U(1) charge $q$ transforms as $\psi
\rightarrow \omega^q \,\psi$; $\chi$ is invariant since $\omega^Q=1$.
Localization of two fields, which we suggestively name $\phi_S$ and
$\phi_T$, at branes separated in the extra dimension forbids local
couplings that involve $\phi_S$ and $\phi_T$ together.

The beneficial effects of localization in the scenario just described
can be captured in a 4D theory via dimensional
deconstruction~\cite{decon}. Let us latticize the extra dimension,
assuming $n$ sites with spacing $a$, where the length of the extra
dimension between its orbifold fixed points is given by $L=n\,a$.  The
deconstructed theory has the 4D gauge group U(1)$^n$, with a U(1)
group factor associated with each lattice site.  The symmetry is
spontaneously broken by $n - 1$ link fields $\Sigma_i$, $i=1, \ldots ,
n-1$, that transform in the bifundamental representation $(+1,-1)$
under U(1) factors U(1)$_i$, U(1)$_{i+1}$ associated with neighboring
lattice sites.  The bulk field $\chi$ is mapped to $n$ fields
$\chi_i$, $i=1, \ldots , n$, with one field at each lattice site.  In
the 4D theory, the nonvanishing $\chi$ profile corresponds to $\langle
\chi_i \rangle \neq 0$ and has the property
\begin{equation}
e^{i \, \alpha_i \,Q} \langle \chi_i \rangle = \langle \chi_i \rangle,
\end{equation}
where $\alpha_i \equiv \alpha(x^\mu,y_i)$ for $\alpha_i=2 \pi /Q$ for
$i=1, \ldots , n$, which together build a $(Z_Q)^n$ symmetry.  The
fields $\phi_S$ and $\phi_T$ located at the ends of the
extra-dimensional interval transform under different $Z_Q$ symmetries,
namely $(Z_Q)_1$ and $(Z_Q)_n$.  However, the link fields $\Sigma_i$
transform as $\Sigma_i \rightarrow \omega_i \omega_{i+1}^{-1}
\Sigma_i$ under the factors $(Z_Q)_i$ and $(Z_Q)_{i+1}$ associated
with the $i^{\rm th}$ and $(i+1)^{\rm th}$ lattice sites,
respectively, breaking these factors to the diagonal $(Z_Q)_{i,i+1}$
subgroup.  The collective effect of the $\Sigma_i$ is to spontaneously
break $(Z_Q)^n \rightarrow Z_Q$.  The remaining discrete symmetry can
be broken by assigning the $\phi_S$ or $\phi_T$ fields nontrivial
$Z_Q$ charges and vevs.

The connection to the $A_4 \times A_4$ model presented earlier is
immediate.  For a larger continuous gauged bulk flavor symmetry [for
example, O(3)] and a field $\chi$ whose profile leaves an $A_4$
subgroup invariant, the simplest two-site deconstruction yields a
theory with an $A_4 \times A_4$ discrete symmetry broken to its
diagonal subgroup by a single link field, corresponding to the
bi-triplet $\Sigma$ studied earlier; localization leads to the
triplets $\phi_S$ and $\phi_T$ transforming under different $A_4$
factors.  The model we presented, however, was constructed from a
purely four-dimensional perspective, without any requirement that the
theory correspond exactly to a deconstructed higher-dimensional
theory.  This starting point provided us the extra freedom to
introduce additional symmetries and assign charges as needed.  The
extra freedom facilitated, for example, the construction of an
explicit model of the symmetry-breaking sector, something that is
rarely presented in studies of deconstructed theories.

\section{Conclusions}\label{sec:conc}

In this paper, we have considered a model of TB neutrino mixing based
on $A_4 \times A_4$ symmetry.  Triplet flavon fields $\phi_S$ and
$\phi_T$ of conventional $A_4$ models transform under different $A_4$
factors so that the correct symmetry-breaking pattern in the charged-
and neutral-lepton sectors occurs.  We constructed an explicit
superpotential that leads to the $\phi_S$ and $\phi_T$ vevs as well as
the vevs in a bi-triplet field $\Sigma$ that breaks the $A_4
\times A_4$ symmetry to its diagonal subgroup; the combined effect of
the $\phi_S$, $\phi_T$, and $\Sigma$ vevs is to break the original
flavor symmetry completely.  To eliminate undesired couplings between
the $\Sigma$ and triplet fields, an additional $Z_3$ symmetry was
imposed on the symmetry-breaking sector.  If this $Z_3$ is allowed to
act on the lepton fields, then charge assignments can be found that
lead to an appropriate charged-lepton mass hierarchy.  We explained
how the assignment of the triplet flavons to different $A_4$ factors
mimics the effects of localization in a 5D theory in which the flavor
symmetry is a discrete subgroup of a continuous gauged flavor symmetry
in the bulk.

This analogy suggests an interesting direction for future study,
namely, models with three $A_4$ factors and two bi-triplet fields,
$\Sigma_1 \sim ({\bf 3}, {\bf 3},{\bf 1}_0)$ and $\Sigma_2 \sim ({\bf
1}_0, {\bf 3},{\bf 3})$, that break the symmetry to the diagonal $A_4$
subgroup.  The hierarchy in the charged-lepton masses can be
accommodated if a nonzero $m_\mu$ requires only the breaking of the
first two $A_4$ factors, while a nonzero $m_e$ requires the breaking
of the first and third $A_4$s via the collective symmetry-breaking
effects of the product $\Sigma_1 \Sigma_2$.  A similar approach may be
successful in accommodating quark flavor as well.  The
symmetry-breaking sector of such a model may be more difficult to
construct explicitly than the model presented here, but is worthy of
further study, especially if complete theories of fermion masses can
be constructed without requiring the imposition of additional
symmetries.

\begin{acknowledgments}
\vskip -2.5ex
This work was supported by the National Science Foundation under Grant
Nos.\ PHY-0757481 (C.D.C.) and PHY-0757394 (R.F.L.). We thank Josh Erlich
for useful comments.
\end{acknowledgments}


\end{document}